\documentclass[journal]{IEEEtran}
\usepackage[T1]{fontenc}
\usepackage[utf8]{inputenc}
\usepackage{amsmath,cite,url,amssymb}
\usepackage{graphicx}
\usepackage{color}
\usepackage{comment}
\usepackage{float} 
\usepackage{array} 
\usepackage{xurl}

\newcounter{row}
\newcommand{\nextrow}{\refstepcounter{row}\arabic{row}}

\begin{document}

\title{The Costs of Reproducibility in Music Separation Research: a Replication of Band-Split RNN}
\author{Paul Magron, Romain Serizel, Constance Douwes
\thanks{P. Magron and R. Serizel are with Université de Lorraine, CNRS, Inria, LORIA, F-54000 Nancy, France  (e-mail: paul.magron@inria.fr, romain.serizel@loria.fr). \\Constance Douwes is with Centrale Med, Aix Marseille Univ, CNRS, LIS, Marseille, France (e-mail: constance.douwes@lis-lab.fr).}
}

\maketitle

\begin{abstract}
Music source separation is the task of isolating the instrumental tracks from a music song. Despite its spectacular recent progress, the trend towards more complex architectures and training protocols exacerbates reproducibility issues. The band-split recurrent neural networks (BSRNN) model is promising in this regard, since it yields close to state-of-the-art results on public datasets, and requires reasonable resources for training. Unfortunately, it is not straightforward to reproduce since its full code is not available. In this paper, we attempt to replicate BSRNN as closely as possible to the original paper through extensive experiments, which allows us to conduct a critical reflection on this reproducibility issue. Our contributions are three-fold. First, this study yields several insights on the model design and training pipeline, which sheds light on potential future improvements. In particular, since we were unsuccessful in reproducing the original results, we explore additional variants that ultimately yield an optimized BSRNN model, whose performance largely improves that of the original. Second, we discuss reproducibility issues from both methodological and practical perspectives. We notably underline how substantial time and energy costs could have been saved upon availability of the full pipeline. Third, our code and pre-trained models are released publicly to foster reproducible research. We hope that this study will contribute to spread awareness on the importance of reproducible research in the music separation community, and help promoting more transparent and sustainable practices.
\end{abstract}

\begin{IEEEkeywords}
Music source separation, band-split RNN, replication study, reproducible research, reproducibility crisis.
\end{IEEEkeywords}


\section{Introduction}
\label{sec:intro}

Music source separation (MSS)~\cite{Cano2019} aims to extract the instrumental tracks that add up to form a music song. This finds application in, e.g., automatic remixing for users with cochlear implants~\cite{Pons2016, Tahmasebi2020}, karaoke/backing track generation~\cite{Tachibana2016, Rafii2018}, or downstream music information retrieval tasks~\cite{Ono2010, Lin2021}.

Thanks to the advent of deep learning, MSS has experienced remarkable performance improvements over the past decade. A wealth of models, architectures, and optimization strategies have been proposed, with a trend towards deeper and more intricate networks. Early approaches focused on spectrogram-domain modeling~\cite{Stoter2019umx, Hennequin2020spleeter}, while more recent models rely on complex-valued frequency-domain~\cite{Luo2023bsrnn,Lu2024} or time/hybrid-domain networks~\cite{Defossez2021hybrid}. Various neural architectures have been leveraged, including recurrent~\cite{Luo2023bsrnn, Drossos2018}, convolutional~\cite{Defossez2021hybrid, Chandna2018}, or transformer~\cite{Rouard2023htdemucs, Lu2024} networks. Community-driven initiatives such as the music demixing challenges (MDX)~\cite{Mitsufuji2022mdx21, Fabbro2024mdx23} have stimulated this quest for performance, resulting in nowadays outstanding separation quality when measured on controlled benchmarks.

Despite this progress, most recent methods suffer from several drawbacks from a \textit{reproducible research} perspective. For instance, winners of the latest MDX challenge~\cite{Fabbro2024mdx23} are bags of models, which can be tedious to train as they depend on multiple base architectures. Other examples are best performing models obtained via fine-tuning on a private dataset, which makes replication plain impossible~\cite{Rouard2023htdemucs}. Lastly, some systems require such a large computing capacity that retraining is prohibitively long when only a fraction of these resources is available~\cite{Lu2024}. These factors exacerbate the reproducibility crisis that plagues many field of applied machine learning~\cite{Dacrema2019,Belz2021}, including music information retrieval~\cite{Six2018,McFee2019}.

Among state-of-the-art MSS methods, band-split recurrent neural network (BSRNN)~\cite{Luo2023bsrnn} appears promising from a reproducibility standpoint. Indeed, it outperforms more recent methods trained and evaluated on the openly available MUSDB18-HQ dataset~\cite{Rafii2019musdb18hq}, thus it is a strong baseline that allows for fair comparison with competing techniques. It is also a single architecture, which is easier to train and fine-tune than bags of models, and the reported training time and hardware are rather reasonable compared to more recent approaches~\cite{Lu2024}. Unfortunately, to the best of our knowledge there is no available implementation of the \emph{full pipeline} that allows us to reproduce the results reported in the paper, including the code for data preparation, and detailed training and evaluation scripts. Besides, while unofficial implementations contain useful resources, their performance is substantially lower than that of the original paper (see Section~\ref{sec:motivation_issues}).

This work addresses the topic of reproducibility in the field of MSS research~\cite{McFee2019,Sculley2015}. Even though replication studies are less common in applied machine learning than in other domains such as pharmacology~\cite{Ioannidis2013pharma} or ecology~\cite{Johnson2002wildlife}, it remains important to take a step back and reflect on previous work, in order to assess progress at the community level. To that end, we propose to replicate BSRNN as closely as possible to the original paper~\cite{Luo2023bsrnn}. In particular, we describe the training and evaluation protocol in details, in order to outline difficulties encountered while replicating the work. We conduct extensive experiments to study various parameters, design choices, and training strategies. Since we were unsuccessful in reaching the performance reported in the original paper, we implement several variants that could potentially bridge this gap. This ultimately yields an \textit{optimized} BSRNN model whose performance is superior to the paper's results. This in-depth investigation allows us to conduct a critical reflection on this reproducibility issue and its various \emph{costs}. To summarize, the main contributions of our work are as follows:
\begin{enumerate}
    \item We conduct extensive experiments for evaluating BSRNN base architecture parameters and its variants. This yields several insights on the model design and training pipeline, and sheds light on potential future improvements for MSS. Besides, even though this work's primary focus is not to further the state-of-the-art, our experiments ultimately yield optimized BSRNN variants that largely outperform the original paper's results.
    \item We show via in-depth discussion that MSS research could be both faster and more energy-effective upon favoring reproducibility. We put a particular emphasis on the energy footprint of this project, as it is essential to encourage more sustainable practices in all fields of machine learning\cite{Henderson2020, Strubell2020}, including the music information retrieval community~\cite{Holzapfel2024greenmir}.
    \item To comply with this objective and to foster reproducible research, we publicly release our code\footnote{\url{github.com/magronp/bsrnn}} and pre-trained models.\footnote{\url{zenodo.org/records/13903584}} These can be useful to the community as they are light and open alternatives to the best performing MSS system~\cite{Lu2024}.
\end{enumerate}

The rest of this paper is structured as follows. Section~\ref{sec:motivation} presents an overview of recent MSS methods, outline their reproducibility issues, and motivates this replication study. Section~\ref{sec:model} describes the BSRNN model, as well as some architecture variants we investigated. Section~\ref{sec:protocol} presents the experimental protocol, and Section~\ref{sec:results} details the results. Section~\ref{sec:discussion} subsequently discusses the costs of reproducibility. Finally, Section~\ref{sec:conclu} draws some concluding remarks.

\section{Overview and motivation}
\label{sec:motivation}

This first section aims at motivating conducting such a replication study. To that end, we outline reproducibility issues with state-of-the-art MSS systems, whose performance are reported in Table~\ref{tab:sdr_comparison}. In order to keep this section brief, we describe the dataset and evaluation metric in~Section~\ref{sec:protocol}, and we detail the process for collecting this data in Appendix~\ref{sec:appx:perfreport}.

\subsection{Model comparison}
\label{sec:motivation_comparison}

\begin{table}[t]
	\center
    \caption{Separation performance (chunk SDR in dB) for state-of-the-art MSS models on the MUSDB18-HQ test set.}
	\label{tab:sdr_comparison}
	\begin{tabular}{m{3.2cm}m{0.6cm}m{0.6cm}m{0.6cm}m{0.6cm}m{0.8cm}}
    \hline
    \hline
         & Vocals & Bass & Drums & Other & Average \\
		\hline \\
		Training on MUSDB18-HQ & & & & & \\
		\hspace{0.3em} CWS-PResUNet~\cite{Liu2021presunet}  & 8.92 & 5.93 & 6.38 & 5.84 & 6.77 \\
		\hspace{0.3em} KUIELab-MDX-Net~\cite{Kim2021kuielab} &  8.97 & 7.83 & 7.20 & 5.90 & 7.47 \\
		\hspace{0.3em} Hybrid Demucs~\cite{Defossez2021hybrid} & 8.35 & 8.43 & 8.12  & 5.65 & 7.64  \\
        \hspace{0.3em} HT Demucs~\cite{Rouard2023htdemucs}  & 7.93 & 8.48  & 7.94 & 5.72 & 7.52  \\
        \hspace{0.3em} BSRNN~\cite{Luo2023bsrnn} & 10.01  & 7.22 & 9.01  & 6.70 & 8.24  \\
        \hspace{0.3em} SIMO-BSRNN~\cite{Luo2024simo} & 9.73 & 7.80 & 10.06 & 6.56 & 8.54 \\
        \hspace{0.3em} BS-RoFormer~\cite{Lu2024} & 10.66  & 11.31  & 9.49 & 7.73 & 9.80   \\
        \hspace{0.3em} DTTNet~\cite{Chen2024} & 10.12  & 7.45  & 7.74 & 6.92 & 8.06 \\
        \hspace{0.3em} TFC-TDF UNet v3~\cite{Kim2023tfctdf} &  9.59 & 8.45  & 8.44 & 6.86  & 8.34 \\
        \hline \\
        Training with extra data & & & & & \\
        \hspace{0.3em} Hybrid Demucs~\cite{Defossez2021hybrid} & 8.75 & 9.13  & 9.31  & 6.18 & 8.34  \\
        \hspace{0.3em} HT Demucs~\cite{Rouard2023htdemucs}  & 9.37 & 10.47  & 10.83  & 6.41 & 9.27  \\
        \hspace{0.3em} BSRNN~\cite{Luo2023bsrnn} & 10.47  & 8.16 & 10.15  & 7.08 & 8.97  \\
        \hspace{0.3em} BS-RoFormer~\cite{Lu2024} & 12.72  & 13.32  & 12.91 & 9.01 & 11.99   \\
       \hline
       \hline
	\end{tabular}
\end{table}

Let us focus on the upper part of Table~\ref{tab:sdr_comparison}, which reports results using the openly available MUSDB18-HQ dataset only. While we indicate CWS-PResUNet's~\cite{Liu2021presunet} performance for reference, it is significantly lower than that of more recent models, thus we do not consider it for a replication study. KUIELab-MDX-Net~\cite{Kim2021kuielab} and the reported Hybrid Demucs~\cite{Defossez2021hybrid} are ensemble / bags of models, which means they combine the outputs of several sub-models. These sub-models can either rely on different base architectures, or instances of the same architecture tuned and trained with different hyperparameters. As a result, they are particularly tedious to optimize, which we illustrate by quoting the authors of Hybrid Demucs:

\begin{quote}
``While suitable for a competition, such a complex model does get in the way of reproducing easily the performance achieved.''~\cite[p.10]{Defossez2021hybrid}
\end{quote}
Furthermore, while HT Demucs~\cite{Rouard2023htdemucs} is more advanced than Hybrid Demucs, its performance is inferior when not trained using extra data. This outlines than while also relevant for competitions or for a private usage, HT Demucs might not be a relevant comparison reference for research papers.

DTTNet~\cite{Chen2024}, SIMO-BSRNN~\cite{Luo2024simo}, and BS-RoFormer~\cite{Lu2024} are variants that build upon the BSRNN model. On the one hand, DTTNet is a lightweight version that yields a slightly lower performance than BSRNN. On the other hand, BS-RoFormer incorporates transformers within BSRNN, which, despite yielding the best performance, becomes prohibitively large for training with our computing resources.\footnote{Each BS-RoFormer model is trained for 4 weeks using 16 Nvidia A100 (80 GB) GPUs in the corresponding paper~\cite{Lu2024}. Adapting this setup to our largest cluster (see Section~\ref{sec:protocol-optim}) would require to divide the global batch size by at least a factor of 4. As a result, a single training run would take about 4 months per instrument (of which there are four in the dataset).} Lastly, SIMO-BSRNN refines the architecture and training process of BSRNN, but it shares BSRNN's reproducibility drawbacks. Therefore, BSRNN appears to be the most suitable model for replication among these variants.

Finally, TFC-TDF UNet v3~\cite{Kim2023tfctdf} performs similarly to BSRNN, but the latter had a larger impact in the MSS community. Indeed, BSRNN has fuelled future developments (e.g., the above-mentioned variants), and was exploited for alternative tasks such as cinematic separation~\cite{Watcharasupat2023} or speech enhancement~\cite{Yu2023sebsrnn}. Therefore, BSRNN appears as the best candidate for replication among these state-of-the-art MSS models.

Note that the lower part of the table displays results obtained with training and/or fine-tuned using extra (private) data. These are solely reported for providing insight about how much gain can be made by fine-tuning a specific model (e.g., about 2.2 dB for BS-RoFormer), but these results are by design nonreplicable. Besides, they allow neither to assess one model's best / maximum potential nor to perform actual comparison between models, since these extra private datasets are different from one paper to another.

\subsection{Problem setting}
\label{sec:motivation_issues}

Let us now address reproducibility issues that are specific to BSRNN. The authors have shared some code for the model definition in the context of the MDX challenge.\footnote{\url{gitlab.aicrowd.com/Tomasyu/sdx-2023-music-demixing-track-starter-kit}} However, there is no available official implementation of the \emph{full pipeline} that allows us to reproduce the results reported in the paper. This includes the code for data preprocessing, detailed training scripts with optimizer and scheduler parameters, and evaluation functions including the inference procedure for processing whole songs. Yet, these are of paramount importance for reproducing any deep learning-based system's performance. Besides, even the afore-mentioned released piece of code cannot be used readily, as it does not make it possible to compute the loss according to the original paper~\cite{Luo2023bsrnn} (see Section~\ref{sec:protocol-loss}). The same applies to SIMO-BSRNN, for which, to the best of our knowledge, no official model definition is provided, while additional implementation details are needed.

Thanks to its popularity in the research community, unofficial BSRNN implementations exist, but they do not fully alleviate the afore-mentioned problems. For instance, A. Amatov's implementation\footnote{\url{github.com/amanteur/BandSplitRNN-PyTorch}} achieves a substantially lower performance to that of the original ($6.67$ vs. $10.01$ dB for the \texttt{vocals} track). C. Landschoot adapted this implementation to a different MSS dataset,\footnote{\url{github.com/crlandsc/Music-Demixing-with-Band-Split-RNN}} and another project\footnote{\url{github.com/sungwon23/BSRNN}} leverages BSRNN for speech enhancement~\cite{Yu2023sebsrnn}, therefore the corresponding pipelines are not appropriate for the original task.

As a result, there is currently no public implementation that allows one to reproduce the results from the original BSRNN paper. This motivates us to conduct a replication study of BSRNN, as well as releasing our code as a fully standalone implementation along with trained models, in order to foster further research or development.

\section{Model and variants}
\label{sec:model}

\begin{figure*}[t]
    \centering
    \includegraphics[width=0.85\linewidth]{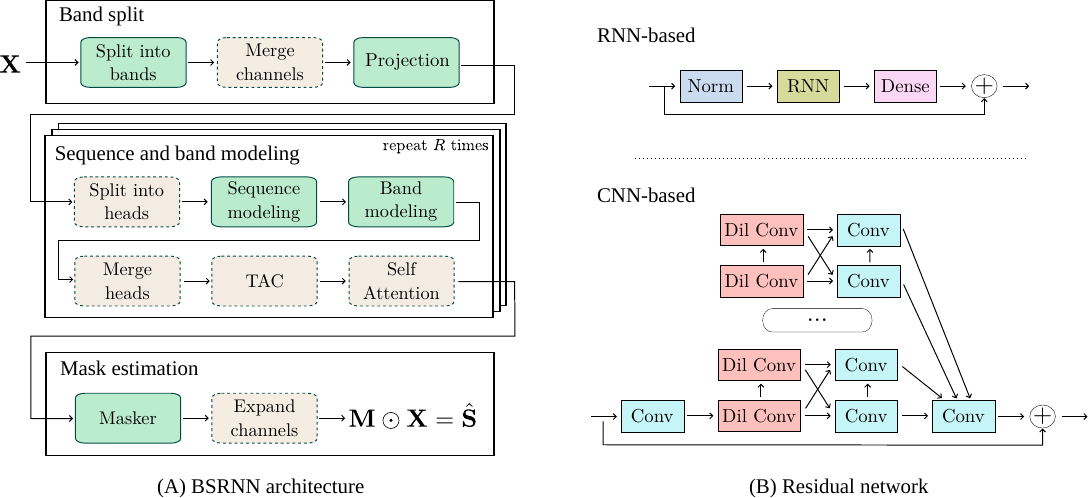}
    \vspace{-0.5em}
    \caption{Overview of BSRNN and its variants (A). Blocs with dashed line contours denote variants not in the original architecture. The residual network (B) in the sequence / band modules are based on either RNNs, as in the original model, or dilated convolutions.}
    \label{fig:bsrnn}
    \vspace{-1em}
\end{figure*}

In this section, we briefly present the original BSRNN model. Then, in an attempt to bridge the performance gap between our implementation's and the paper's results, we propose a few architecture variants. The original model and its variants are illustrated in Figure~\ref{fig:bsrnn}.

\subsection{Original BSRNN architecture}
\label{sec:model-orig}

BSRNN takes as input the complex-valued short-time Fourier transform (STFT) of the mixture $\mathbf{X} \in \mathbb{C}^{F \times T}$, where $F$ and $T$ denote the number of frequency bands and time frames, respectively, and predict the STFT of a target source $\mathbf{S} \in \mathbb{C}^{F \times T}$. In practice, complex-valued STFTs are represented as stacks of their real and imaginary parts, since such a real-valued tensor can easily be processed with a neural network. BSRNN consists of the three following modules.

First, the \textit{band split} module decomposes the input STFT into~$K$ subband spectrograms with variable bandwidth.
These are subsequently projected into a deep latent space with dimension~$N$, and the obtained deep subband features are then stacked back into a fullband tensor with dimensions $N \times K \times T$. The band split scheme is designed and optimized for each source specifically, which yields instrument-specific models of different size.

Then, the \textit{sequence and band modeling} module applies two residual networks (one after the other), whose basic structure consists of a group normalization, a bidirectional long short-term memory (LSTM), and a dense layer. The sequence and band RNN act across the $T$ time frames and $K$ bands, respectively. This dual-path approach~\cite{Luo2020dprnn} optimally exploits the structure of audio/music signals by capturing dependencies across both time and frequencies. Such RNNs are stacked $R$ times to form a deeper network.

Finally, the \textit{mask estimation} module splits the output of the previous module into subbands, and each subband feature is fed to a multi-layer perceptron (MLP) that predicts a subband mask. The MLPs use a hidden size~$\mu \times N$, where $\mu=4$. Subband masks are then assembled into a fullband mask ${\mathbf{M} \in \mathbb{C}^{F \times T}}$ that is multiplied with the input STFT to yield the source estimate: $\hat{\mathbf{S}} = \mathbf{M} \odot \mathbf{X}$, which is finally reverted to time-domain via inverse STFT (iSTFT).

\subsection{Variants}
\label{sec:model-variants}

Here we present our proposed architecture variants, focusing on keys concepts for a clarity purpose. We refer the interested reader to our code for full implementation details.

\subsubsection{Stereo modeling}
\label{sec:model-stereo}

By design, BSRNN is a single-channel model that processes the two channels of the (stereo) input mixture independently, as if they were two single-channel streams coming from different music songs. Not accounting for the cross-channel information might limit the system's performance, and processing the left and right channels independently increases the computational burden unnecessarily.

To alleviate this issue, we propose a stereo variant of BSRNN that is based on simple modifications of the band split and mask estimation modules. We view a stereo input mixture as a 2-channel STFT $\mathbf{X} \in \mathbb{C}^{2 \times F \times T}$, which is first split into subbands. Then, the left and right channels of each subband spectrogram are concatenated and jointly projected into a deep subband feature with dimension $N$. The sequence and band modeling module is identical, and the masker's MLP has twice as many outputs, corresponding to the two channels of the estimated source. In what follows we refer to this strategy as the naive stereo model. It corresponds to the ``Merge channels'' and ``Expand channels'' blocs in Figure~\ref{fig:bsrnn}-(A).

Alternatively, we propose to leverage a transform-average-concatenate (TAC) module, originally tailored for speech separation~\cite{Luo2020tac}, and recently used for MSS in the SIMO-BSRNN variant~\cite{Luo2024simo}. In a nutshell, the TAC module first \emph{transforms} a multichannel input representation via a dense layer; then applies \emph{averaging} over channels and a second dense layer; and finally \emph{concatenates} this second layer's output with the first layer's transformed representation, before passing it to a third dense layer. Since this TAC module shares information across channels, it enables stereo-aware modeling. We incorporate it after sequence and band modeling~\cite{Luo2024simo}, and the dense layers' activation function is either a hyperbolic tangent (TanH), as in SIMO-BSRNN~\cite{Luo2024simo}, or a parametric rectified linear unit (PReLU), as in the original TAC paper~\cite{Luo2020tac}.

\subsubsection{Alternative layers}
\label{sec:model-bscnn}

We investigate alternative layers to LSTMs in the sequence and band modeling module. Preliminary experiments showed that gated recurrent units yield similar results to LSTMs (as outlined in other studies~\cite{Watcharasupat2023}), and replacing each RNN with a simple one-layer convolutional neural network (CNN) induces a large performance drop, although these are much faster to train because of reduced memory constraints. However, this approach is quite naive, as such CNNs cannot properly model sequential data with long-term dependencies. Therefore, we propose to use a stack of dilated CNNs, since they have shown promising for source separation, and are competitive with RNNs due to their increased receptive field~\cite{Defossez2021hybrid, wang23ca_interspeech}. We implement the architecture from Wang~\cite{wang23ca_interspeech}, which yields a fully-convolutional model denoted band-split CNN (BSCNN) and illustrated in Figure~\ref{fig:bsrnn}-(B).

\subsubsection{Self-attention}
\label{sec:model-attention}

The self-attention mechanism has shown promising in many tasks involving sequential data such as speech enhancement and separation~\cite{Pandey2021, Wang2023tfgridnet}. In particular, it was used in TFGridNet~\cite{Wang2023tfgridnet}, a model that shares similarities with BSRNN, as it projects learns dependencies across bands and time frames via a dual-path-like architecture. In TFGridNet, the self-attention mechanism was observed to yield a noticeable performance boost, with a negligible impact in terms of model size and computational burden. We incorporate $N_a$ attention heads with an encoding dimension $E_a$ within each sequence and band modeling module, using the implementation from the ESPnet toolbox~\cite{Yan2023espnet}.

\subsubsection{Multi-head mechanism}
\label{sec:model-multihead}

Chen et al.~\cite{Chen2024} proposed to replace the sequence and band modeling modules with so-called ``improved'' modules. These boil down to splitting the input feature of dimensions $ N \times K \times T$ into $H$ heads, each with dimensions $H \times N/H \times K \times T$. The RNNs then operate in parallel across heads using a smaller hidden dimension of $N/H$ instead of $N$, which drastically reduces the number of (redundant) parameters. This corresponds to the ``Split into heads'' and ``Merge heads'' blocs in Figure~\ref{fig:bsrnn}-(A).

\section{Experimental protocol}
\label{sec:protocol}

This section describes our experiment protocol. Since it is largely similar to that of the original BSRNN paper~\cite{Luo2023bsrnn}, we mostly focus on aspects for which variants are considered, or details that need special care and clarification.

\subsection{Dataset and data processing}
\label{sec:protocol-data} 

We use the openly available MUSDB18-HQ dataset~\cite{Rafii2019musdb18hq} for all experiments. It consists of 150 stereo songs sampled at 44100~Hz, with pairs of mixtures and corresponding four isolated sources: \texttt{vocals}, \texttt{bass}, \texttt{drums}, and \texttt{other}. The songs are split into 86, 14, and 50 tracks for training, validation, and testing, respectively. The STFT is computed using a 2048 sample-long Hanning window with a hop size of 512.

We generate 20,000 training samples per epoch following the original paper's methodology. First, we preprocess training tracks with a source activity detector (SAD) that removes the silent regions. Then, training samples are generated on-the-fly as follows:
\begin{itemize}
    \item[(i)] We randomly select sources from different songs.\footnote{Preliminary experiments did not reveal any difference between shuffling tracks only once when training starts, or re-shuffling tracks at each epoch. We use the former as it is slightly faster. Besides, note that even though this process yields inconsistent music mixtures - a phenomenon termed \emph{cacophony} - it was shown to effectively improve MSS performance~\cite{Jeon2024}.}
    \item[(ii)] We extract random 3~s-long chunks for each source.
    \item[(iii)] We apply a random scale factor to each chunk such that its energy is in a $[-10, 10]$ dB range.
    \item[(iv)] We drop each chunk with probability~0.1 to simulate silent sources.
    \item[(v)] We add chunks to form mixtures.
\end{itemize}
We also consider an alternative training data generation process, without SAD preprocessing nor random chunk dropping, that uses the same augmentations as in Open-Unmix (UMX)~\cite{Stoter2019umx}, i.e., randomly swapping the channels with a probability of~$0.5$, and rescaling the chunks' energy using a linear gain between~0.25 and 1.25.

\subsection{Model configuration}
\label{sec:protocol-config}

In order to accelerate prototyping, most experiments use a small model, i.e., $N=64$ and $R=8$, while the original paper uses $N=128$ and $R=12$, which herein corresponds to the models denoted \textit{large} or \textit{optimized}. All other hyper-parameters are chosen as per the original paper~\cite{Luo2023bsrnn}, unless specified explicitly.

\subsection{Metrics}
\label{sec:protocol-sdr}

We assess source separation quality in terms of signal-to-distortion ratio (SDR) expressed in dB (higher is better)~\cite{Vincent2006}. We consider the following two variants of the SDR:
\begin{itemize}
    \item The \textit{utterance} SDR (uSDR) is equal to a basic signal-to-noise ratio computed on whole songs. We report the mean across songs, as in the MDX challenges~\cite{Mitsufuji2022mdx21,Fabbro2024mdx23}, where it is used as evaluation metric.
    \item The \textit{chunk} SDR (cSDR) is computed by taking the median SDR across 1 s-long chunks. We report the median across songs, as per the signal separation evaluation campaign (SiSEC) guideline~\cite{Stoter2018}, and we compute the cSDR using the museval toolbox~\cite{Stoter2021museval}.
\end{itemize}
%

\subsection{Training}
\label{sec:protocol-training}

\subsubsection{Loss}
\label{sec:protocol-loss}

BSRNN is trained using a combination loss ${\mathcal{L} = \mathcal{L}_{\text{time}} + \mathcal{L}_{\text{STFT}}}$ that consists of a time-domain term:
\begin{equation}
    \mathcal{L}_{\text{time}} = | \text{iSTFT}(\mathbf{S}) - \text{iSTFT}(\hat{\mathbf{S}}) |_1,
\end{equation}
and an STFT-domain term:
\begin{equation}
    \mathcal{L}_{\text{STFT}} = | \mathbf{S}_r - \hat{\mathbf{S}}_r |_1 +  | \mathbf{S}_i - \hat{\mathbf{S}}_i |_1,
    \label{eq:ltf}
\end{equation}
where $|.|_1$ denotes the $\ell_1$ norm, and the subscripts $r$ and $i$ respectively denote the real and imaginary parts.

\subsubsection{Optimizer and hardware}
\label{sec:protocol-optim}

In the original paper, the model is trained using the Adam algorithm with an initial learning rate $\lambda=10^{-3}$, a batch size of 2, and 8 GPUs in parallel, yielding a \textit{global} batch size $B=16$. Unfortunately, we do not have access to enough (large) GPUs to obtain the same global batch size. As a result, we resort to adjusting the learning rate in order to preserve the same \textit{effective} learning rate $\lambda / B$, which ensures similar gradient descent steps~\cite{Goyal2017}. The learning rate is decayed by~$0.98$ every two epochs, and gradient clipping by a maximum gradient norm of 5 is applied. 

Small models are trained using 4 Nvidia RTX 2080 Ti (11~GB) GPUs, except for the \texttt{drums} model with self-attention, which is trained using 4 Nvidia Tesla T4 (15~GB) GPUs because of memory constraints. Large models are trained using 2 Nvidia Tesla L40S (45~GB) GPUs.

\subsubsection{Monitoring}
\label{sec:protocol-monitor}

Training is conducted with a maximum of~$100$ epochs in the original paper~\cite{Luo2023bsrnn}, which we did not observe to be enough for convergence in most cases, thus, we set this number at~$200$.
Besides, the authors apply early stopping ``when the best validation is not found in 10 consecutive epochs''~\cite[IV-A]{Luo2023bsrnn}, which does not clearly state which quantity (loss or SDR) is monitored. By default we chose to monitor the uSDR since computing the cSDR is time-consuming, and we keep a patience of 10 epochs.

\subsection{Inference of whole songs for evaluation}
\label{sec:protocol-inference}

At the evaluation stage (i.e., validation or test), songs cannot be processed entirely at once by the model because of memory constraints. Consequently, they are divided into small (overlapping) segments that are fed to the separation model, and then the estimated chunks are assembled using an overlap-add (OLA) strategy to recover whole sources.

In the BSRNN paper, the evaluation segments are 3~s-long with a 0.5~s hop size, but the exact OLA/reconstruction technique (i.e., windowing, trimming/concatenation, etc.) is not specified. We then implement a simple procedure assuming a rectangular window whose scale factor is set to ensure perfect reconstruction. Alternatively, we split the songs into segments of 10~s with 10\% overlap, and the estimated chunks are assembled using a linear fader to smooth their edges~\cite{Defossez2021hybrid}.

Preliminary experiments showed that the inference procedure has no impact in terms of model selection at the \emph{validation} stage, since the best selected model is the same regardless of mild absolute differences in terms of validation uSDR. As a result, our experiments use the linear fader for validation since it is much faster. However, the inference procedure has an impact onto \emph{test} scores: indeed, differences up to $0.3$ dB for the \texttt{vocals} track were observed for BSRNN~\cite{Luo2023bsrnn}, and similarly for TFC-TDF UNet v3~\cite{Kim2023tfctdf}. While this aspect is often overlooked when comparing models, we will discuss it in Section~\ref{sec:results-test}.


\subsection{Monitoring energy consumption}
\label{sec:protocol-energy}

We monitor the energy consumption of training the various models considered in this work. To that end, we track the energy (measured in kWh) during training via the codecarbon toolbox~\cite{Courty2024codecarbon}. Since this method tends to underestimate energy, we also follow the methodology of the Green Algorithms online calculator~\cite{lannelongue2021green}, which approximates energy consumption based on hardware specifications (we consider a 3~W power per 8~GB of memory). This method has been shown to yield results that are closer to actual power meter readings compared to codecarbon~\cite{jay2023experimental}. We use a power usage effectiveness factor of 1.5, according to our computing platform's recommendation.\footnote{\url{https://intranet.grid5000.fr/stats/indicators}} Note that, as described in Section~\ref{sec:protocol-training}, our various models are trained using different GPUs, which has an impact on energy consumption and therefore might affect comparison~\cite{Serizel2023}. Nevertheless, we use this setup as it allows us to optimally exploit our available hardware.

We also estimate the total consumed energy for the project, including all other models variants (not reported in this paper), preliminary and additional experiments, prototyping, and inference. For simplicity, we report the total project energy computed with the Green Algorithms calculator.

\section{Results}
\label{sec:results}

This section describes our experimental results. We do not report all conducted experiments, as this paper would be prohibitively long. More details can be found in a supporting document,\footnote{\url{github.com/magronp/bsrnn/blob/main/docs/analysis.md}} and we encourage the interested reader to experiment with our code for conducting further investigation.

\subsection{Validation results}

\begin{table*}[t]
    \center
    \caption{Comparison of various models' performance: best model's validation uSDR (in dB), total number of parameters (in millions, ``-'' denotes the same value as in the preceding line), and estimated energy for training all sources (in kWh).}
	\label{tab:val}
	\begin{tabular}{cl|ccccc|ccc}
    \hline
    \hline
                                    &    & \multicolumn{5}{c}{Validation uSDR (dB)} & \# Parameters (M)   &  \multicolumn{2}{c}{Energy (kWh)}  \\
                                    &    & Vocals & Bass   & Drums & Other  & Average  \vspace{0.5em}&  & codecarbon & green algo. \\
    \nextrow\label{tabl:base}   &  Base model: $N=64$, $R=8$                      & $7.7$  & $6.1$  & $9.7$  & $4.8$  & $7.1$  & 32.3   & 127 & 168\\ 
    \hline \\ 
    & \hspace{-2em} \vspace{0.5em} Training parameters & & & & & & &  \\
    \nextrow\label{tabl:acc}    & Accumulating gradients          & $8.0$  & $5.8$  & $9.6$  & $4.9$  & $7.1$  & -      & 129 & 170\\
    \nextrow\label{tabl:monit}  & Monitoring with the loss        & $7.5$  & $6.4$  & $9.3$  & $4.8$  & $7.1$  & -      & 120 & 159\\
    \nextrow\label{tabl:losst}  & Loss domain:  time              & $7.9$  & $6.1$  & $9.4$  & $4.9$  & $7.1$  & -      & 116 & 153 \\
    \nextrow\label{tabl:losstf} & Loss domain:  STFT              & $7.9$  & $6.4$  & $9.6$  & $4.9$  & $7.2$  & -      & 131 & 173 \\
    \hline  \\ 
    & \hspace{-2em} \vspace{0.5em} Original architecture parameters & & & & & & &  \\
    \nextrow\label{tabl:stft}   & STFT: window=4096, hop=1024   & $7.3$  & $5.9$  & $8.7$  & $4.4$  & $6.6$  & 37.1   & 58 & 92 \\
    \nextrow\label{tabl:mask}   & Masker factor $\mu=2$           & $7.9$  & $6.8$  & $9.4$  & $4.4$  & $7.1$  & 20.6   & 110 & 151\\
    \nextrow\label{tabl:large}  & Large model: $N=128$, $R=12$    & $9.2$  & $7.3$  & $10.3$ & $5.8$  & $8.2$  & 146.7  & 230 & 321 \\
    \nextrow\label{tabl:large30}& Large model with patience=30     & $9.5$  & $7.8$  & $10.3$ & $6.3$  & $8.4$  & -      & 354  & 495 \\
    \hline  \\ 
    & \hspace{-2em} \vspace{0.5em} Stereo models & & & & & & & &  \\
    \nextrow\label{tabl:stereo} & Naive                   & $7.7$  & $6.6$  & $8.4$  & $4.0$  & $6.7$  & 37.1   & 78 & 122  \\
    \nextrow\label{tabl:stereo8}& Naive, with $\mu=8$          & $7.9$  & $6.1$  & $8.7$  & $4.3$  & $6.7$  & 81.1   & 87 & 140 \\
    \nextrow\label{tabl:tactanh} & TAC with TanH activation         & $7.6$  & $6.0$  & $9.6$  & $4.3$  & $6.8$  & 34.7   & 117 &  154 \\
    \nextrow\label{tabl:tacprelu}& TAC with PReLU activation       & $7.9$  & $6.5$  & $10.0$  & $4.7$  & $7.3$  & 34.7   & 128 &  167\\
    \hline  \\ 
    & \hspace{-2em} \vspace{0.5em} Architecture variants & & & & & & & &  \\
    \nextrow\label{tabl:bscnn}  & BSCNN                          & $7.3$  & $5.9$  & $9.0$  & $4.2$  & $6.6$  & 29.7   & 113 & 153 \\
    \nextrow\label{tabl:att1}   & Attention: $N_a=1$, $E_a=8$    & $7.7$  & $7.4$  & $10.4$ & $4.8$  & $7.6$  & 33.0   & 151 & 199 \\
    \nextrow\label{tabl:att2}   & Attention: $N_a=2$, $E_a=16$   & $8.2$  & $7.7$  & $10.4$ & $4.9$  & $7.8$  & 33.2   & 157 & 224\\
    \nextrow\label{tabl:multi} & Multi-head module, $H=2$      & $7.6$  & $5.5$  & $9.1$  & $4.0$  & $6.6$  & 22.0   & 91  &  137 \\
    \hline  \\ 
    & \hspace{-2em} \vspace{0.5em} Data generation & & & & & & & & \\
    \nextrow\label{tabl:silent} & Silent target (instead of all sources)& $7.9$  & $6.6$  & $9.5$  & $4.4$  & $7.1$  & 32.3   & 110 & 146 \\
    \nextrow\label{tabl:nosad}  & No SAD; UMX-like augmentations & $8.2$  & $6.9$  & $9.5$  & $5.3$  & $7.5$  & -      & 135 & 179\\
    \hline  \\ 
    & \hspace{-2em} \vspace{0.5em} Optimized models & & & & & & & & \\
    \nextrow\label{tabl:opt}    & Attention, patience = 30, no SAD  & $10.1$ & $9.1$  & $10.9$ & $6.7$  & $9.2$  & 149.9  & 426 & 593 \\
    \nextrow\label{tabl:opttac}    & + TAC & $10.2$ & $10.2$  & $11.3$ & $6.9$  & $9.6$  & 164.1  & 508 & 711 \\
       \hline
       \hline
	\end{tabular}
\end{table*}

We report the best model's validation uSDR and training energy for all variants in Table~\ref{tab:val}. Each line describes the difference with the base model (reported at line~\ref{tabl:base}), which serves as a comparison reference for each variant.

\subsubsection{Base model}
\label{sec:results-base}

\begin{figure}
    \centering
    \includegraphics[width=0.99\linewidth]{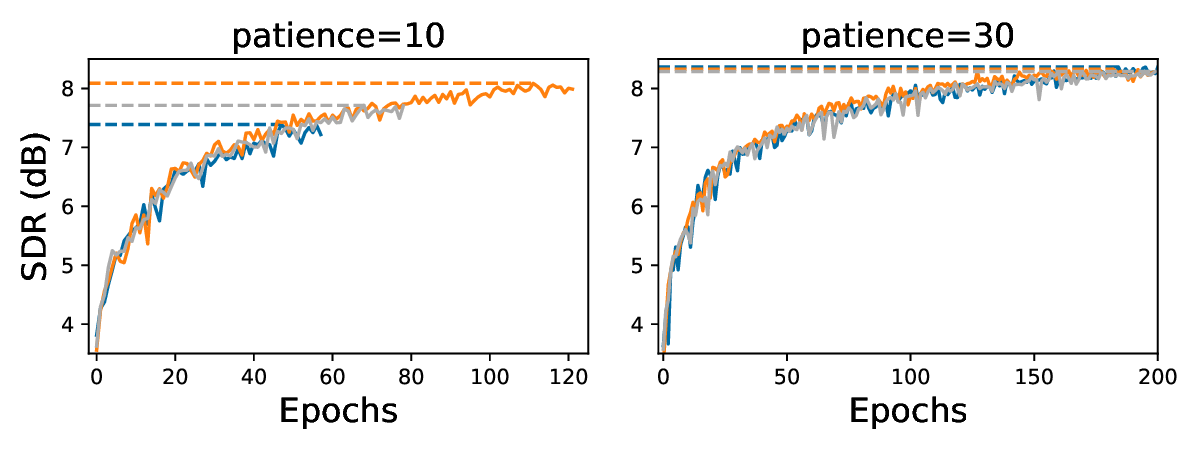}
    \caption{Validation uSDR over epochs for the base model on the \texttt{vocals} track, with a patience of 10 (left) or 30 (right). Each color corresponds to a different run, and the dashed lines correspond to each run's best uSDR.}
    \label{fig:patience}
\end{figure}

Let us first consider a \textit{base model}, which corresponds to the small-size version of BSRNN (see Section~\ref{sec:protocol-config}). As a preliminary experiment, we train it using three different random seeds, and display the validation uSDR over epochs in Figure~\ref{fig:patience} (left) for the \texttt{vocals} track (similar results are obtained for the other sources). We observe that all runs exhibit the same trend, but they yield some variance in terms of best uSDR. This variance is explained by instabilities in the uSDR that stem from the different initial random seed~\cite{Bouthillier2021}, which ultimately causes training to stop at different epochs.

To alleviate this issue, we increase the patience parameter at~30, and we display the results in Figure~\ref{fig:patience} (right). We observe that the mean best uSDR is increased, but more importantly, its variance is largely reduced. Increasing patience is therefore effective to continue training and reduce the impact of the random seed onto the best uSDR. However, this comes at the cost of increasing the number of epochs, and consequently the energy, by a factor $2.3$.

Based on these findings, one could either reduce variance by tuning the random seed as any other hyperparameter~\cite{Bethard2022}, or average multiple runs / increase patience as done above. Nevertheless, to keep our experiments time and energy cost-effective and consistent with the original paper, we report a single run's results with a patience of~10, unless specified explicitly. An exception to this is the base model, for which we report the mean value over the three initial runs, since we have conducted these already.

\subsubsection{Training parameters} \hfill

\noindent \textbf{Accumulating gradients} Instead of adjusting the learning rate as described in Section~\ref{sec:protocol-optim}, we can accumulate gradients over steps before performing descent, which artificially increases the global batch size. Results from line~\ref{tabl:acc} show that both strategies perform similarly. In what follows we adjust the learning rate as we observed it to be more stable when training larger models.

\noindent \textbf{Monitoring criterion} While by default early stopping is performed via monitoring the uSDR (see Section~\ref{sec:protocol-monitor}), we can instead monitor the validation loss. Both approaches yield similar results (\textit{cf}. line~\ref{tabl:monit}), but we chose the former as it allows training to continue for more epochs, which is beneficial in ensuring convergence.

\noindent \textbf{Loss} We conduct training by minimizing each term of the loss individually (see Section~\ref{sec:protocol-loss}). Interestingly, all loss domains (\textit{cf}. lines~\ref{tabl:base},~\ref{tabl:losst}, and~\ref{tabl:losstf}) yield similar results, which contrasts with previous studies that outlined the importance of time-domain training~\cite{Heitkaemper2020}. One explanation is that the STFT-domain loss~\eqref{eq:ltf} treats both the real \textit{and} imaginary parts separately. This likely enforces \textit{phase consistency}~\cite{LeRoux2008phase, Wisdom2019} implicitly, which makes it equivalent to a time-domain loss.

\subsubsection{Original architecture parameters} \hfill

\noindent \textbf{STFT parameters} We consider an STFT with a 4096 sample-long window and a hop size of 1024 samples, as this setup is common among MSS models~\cite{Stoter2019umx, Defossez2021hybrid}. As shown at line~\ref{tabl:stft}, this yields poor result, which we attribute to a reduced time resolution that is not compensated by the increased frequency resolution. Indeed, the latter has little impact since the band split scheme and projection in a latent space of fixed dimension occur early in the network. This notably explains why the drop is more pronounced for the \texttt{drums} than for the \texttt{bass} track, since percussive events are localized in time and thus require a refined time resolution to be properly modeled. Consistently, a larger window of 6144 points~\cite{Chen2024} yields worse results.

\noindent \textbf{Masker size} We propose to reduce the MLP masker size by setting $\mu=2$ (vs. $\mu=4$ by default, see Section~\ref{sec:model-orig}). While this negatively affects performance for the \texttt{drums} and \texttt{other} tracks, it substantially improves the \texttt{bass} results, thus yielding a similar average performance (\textit{cf}. line~\ref{tabl:mask}). Further decreasing $\mu$ at 1 degrades performance more importantly.

\noindent \textbf{Large model} Using the original model size (${N=128}$ and ${R=12}$) yields a large uSDR improvement of 1.1~dB on average compared to the base model (line~\ref{tabl:large}). Nonetheless, such a model was not fully converged (except for the \texttt{drums} track), thus we continue training after increasing the patience, which further improves performance by 0.2~dB (line~\ref{tabl:large30}). However, this improvements appears very limited when further considering that this extended training has resulted in an a $54$\% energy increase. Be that as it may, this model's test set performance is still lower than the original results (see Section~\ref{sec:results-test}), which motivates the next experiments.

\subsubsection{Architecture variants} \hfill

\noindent \textbf{Stereo models} Our naive approach to stereo modeling results in a performance decrease, except for the \texttt{bass} track, as attested by the results from line~\ref{tabl:stereo}.
One way to bridge this gap consists in increasing the masker size ($\mu=8$) to compensate for the larger number of outputs, as two channels must be recovered. While it partly mitigates the performance drop for \texttt{drums} and \texttt{other}, it yields worse results for the \texttt{bass} track (\textit{cf}. line~\ref{tabl:stereo8}). Besides, the model becomes prohibitively large for training when further increasing $N$ and $R$ to match the ``large'' setup. Leveraging a TAC module, similarly to the SIMO-BSRNN variant~\cite{Luo2024simo}, also exhibits a performance drop (see line~\ref{tabl:tactanh}). However, replacing the TanH activation with a PReLU, as originally proposed in the TAC paper~\cite{Luo2020tac}, outperforms the base model, with a more substantial improvement for the \texttt{bass} and \texttt{drums} tracks (line~\ref{tabl:tacprelu}).

\noindent \textbf{BSCNN}
We report at line~\ref{tabl:bscnn} the performance of our proposed BSCNN model, after preliminary experiments to tune the convolution hyperparameters (number of dilated layers, kernel sizes, etc.). We observe that although faster to train and slightly lighter, BSCNN is outperformed by the RNN-based network. A refined and source-specific tuning of the convolution parameters (e.g., considering different parameters for band and sequence modeling) could improve performance, which we leave to future investigation.

\noindent \textbf{Self-attention} As attested by lines~\ref{tabl:att1} and~\ref{tabl:att2}, incorporating self-attention heads is beneficial, except for the \texttt{other} track. In particular, such a small-size \texttt{drums} model with attention performs similarly to a large model with no attention (\textit{cf}. line~\ref{tabl:large}). This approach is therefore an interesting alternative to larger and much more computationally demanding models that completely replace RNNs with transformers~\cite{Lu2024}.

\noindent \textbf{Multi-head module} Results from line~\ref{tabl:multi} show that the multi-head module performs worse for most sources, except for the \texttt{vocals} track. Setting $R=4$~\cite{Chen2024} or further increasing $H$ degrades performance more importantly, though the model becomes much lighter.

\subsubsection{Data generation}

Instead of randomly dropping each \textit{chunk} to simulate silent sources when generating data~\cite[IV-A-2]{Luo2023bsrnn}, we only drop the \textit{target} source, which improves performance for the \texttt{bass} track (\textit{cf}. line~\ref{tabl:silent}). Besides, not performing SAD and applying the UMX-like augmentations further improves performance on average, as observed at line~\ref{tabl:nosad}. This suggests that there is room for improvement for our SAD implementation, since this preprocessing is alleged to greatly benefit the separation. Note however that it is not clear to what extent, since its impact is not evaluated in a specific experiment~\cite{Luo2023bsrnn}.

\subsubsection{Optimized models}
\label{sec:results-opt}

We use this alternative data generation process to train a large model ($N=128$ and $R=12$) with self-attention heads ($N_a=2$ and $E_a=16$), with an increased patience of~30. According to the results displayed at line~\ref{tabl:opt}, this model outperforms our previous large model (line~\ref{tabl:large30}), at the cost of a moderate increase in number of parameters. Further incorporating the TAC module (with PReLU activation) for stereo modeling yields an \textit{optimized} model, which benefits from an extra $0.4$ dB performance (line~\ref{tabl:opttac}).

\subsection{Energy consumption}
\label{sec:results-energy}
 
\begin{figure}
    \centering
    \includegraphics[width=0.9\linewidth]{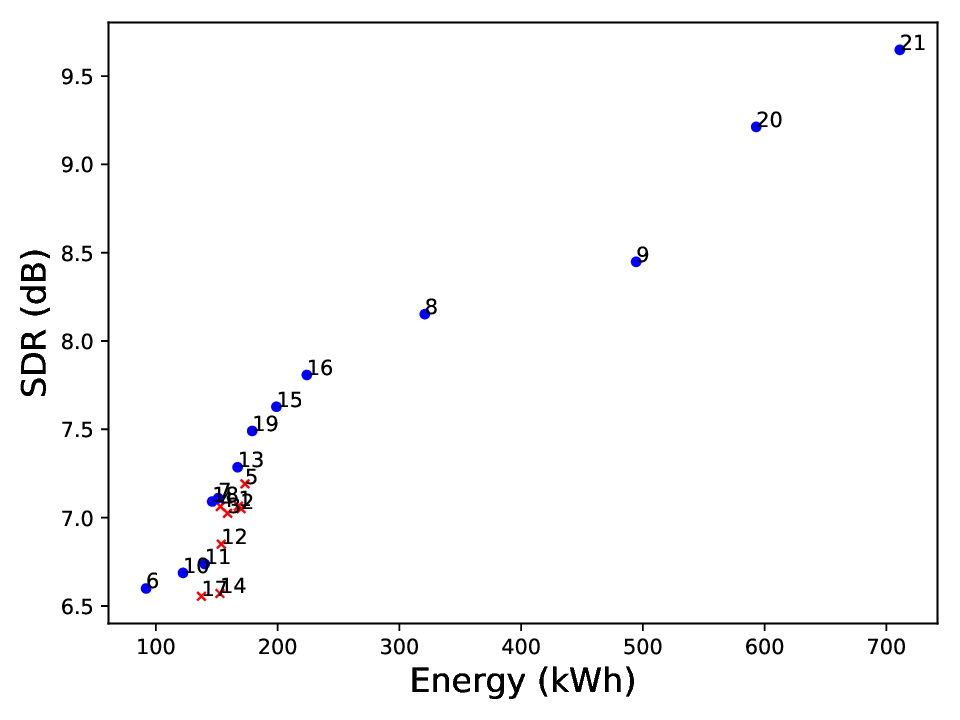}
    \vspace{-0.5em}
    \caption{Performance vs. training energy (estimated using the Green Algorithms calculator) for model variants, where each number marker corresponds to the line number in Table~\ref{tab:val}. Blue dots highlight models that are optimal in a Pareto sense.}
    \label{fig:pareto}
\end{figure}

We report the energy consumed for each model's training in the last columns of Table~\ref{tab:val}. Note that line~\ref{tabl:base} corresponds to the mean over the three runs (see Section~\ref{sec:results-base}). We observe that both energy estimation techniques behave similarly in terms of relative energy variation between models overall. The Green Algorithms calculator yields larger values than codecarbon, which is consistent with previous studies (see \textit{cf}. Section~\ref{sec:protocol-energy}).

Most small-size model variants exhibit a similar energy cost, except for those with lighter architecture / memory requirements, or generally faster convergence (lines~\ref{tabl:stft},~\ref{tabl:stereo}, and~\ref{tabl:multi}), which all reduce energy consumption. Conversely, the larger energy costs of models with attention (lines~\ref{tabl:att1} and~\ref{tabl:att2}) is due to requiring more epochs for reaching convergence. This highlights the importance of considering both performance \textit{and} energy consumption when performing model selection. For instance, while the model at line~\ref{tabl:silent} performs similarly to the base model, it uses less energy for training, which makes it an interesting alternative. We further illustrate this by displaying the performance against the energy in Figure~\ref{fig:pareto}, where we highlight points that are optimal in a Pareto sense~\cite{Douwes2023}, that is, such that there is no other point yielding both a higher SDR and a lower energy. This allows one to discard variants for which both a better performing and less costly alternative exists, such as the model using TAC with TanH activation (line~\ref{tabl:tactanh}) vs. the base model (line~\ref{tabl:base}).

Overall, training all the models discussed in this table amounts to 4,808~kWh, which represents about 7 times the energy cost of the single best model. While this ratio might seem reasonable at first glance, we also estimate the total consumed energy for the project, as described in Section~\ref{sec:protocol-energy}. This amounts to 23~MWh, which is more than 32 times the energy consumption of training the best model, or 136 times that of the base model. To put things in perspective, this is equivalent to the yearly electricity consumption of about 15 persons in Europe.\footnote{\url{https://ec.europa.eu/eurostat/statistics-explained/index.php?title=Electricity_and_heat_statistics}} Such a substantial toll advocates for systematic energy reporting and more sustainable development practices, including more reproducible research, since part of this energy cost could have been saved if a complete and documented code had been available (see Section~\ref{sec:discussion}).

\subsection{Test results}
\label{sec:results-test}

\begin{table*}[t]
    \center
    \caption{Source separation performance on the MUSDB18-HQ test set (uSDR and cSDR, in dB).}
	\label{tab:test_results}
	\begin{tabular}{l|cc|cc|cc|cc|cc}
    \hline
    \hline
         &  \multicolumn{2}{c|}{Vocals} & \multicolumn{2}{c|}{Bass} &  \multicolumn{2}{c|}{Drums} &  \multicolumn{2}{c|}{Other}  &  \multicolumn{2}{c}{Average} \\
		 & uSDR & cSDR & uSDR & cSDR  & uSDR & cSDR & uSDR & cSDR & uSDR & cSDR \\
Paper's results~\cite{Luo2023bsrnn}  & $10.04$ & $10.01$ & $6.80$ & $7.22$  & $8.92$ & $9.01$ & $6.01$  & $6.70$ & $7.94$ & $8.24$ \\
    \hline
    Our implementation      &   &   & &  &  & &  & & & \\
    \hspace{0.5em} Fader                   & $9.16$  & $9.14$  & $6.51$ & $7.72$  & $8.55$ & $8.07$ & $5.44$  & $5.68$ & $7.42$ & $7.65$ \\
   	\hspace{0.5em} OLA, hop = 1.5 s        & $9.10$  & - & $6.47$ & - & $8.33$ & - & $5.44$  & - & $7.34$ & -\\
   	\hspace{0.5em} OLA, hop = 0.5 s        & $9.18$  & - & $6.60$ & - & $8.43$ & - & $5.53$  & - & $7.44$ & -\\
   	\hline
   	
   	 Optimized models  &   &   & &  &  & &  & & & \\
        \hspace{0.5em} oBSRNN            & $9.78$  & $9.81$  & $8.38$ & $9.85$  & $10.28$ & $10.31$ & $5.85$  & $6.31$ & $8.57$ & $9.07$ \\
        \hspace{0.5em} oBSRNN-SIMO       & $10.73$  & $10.66$ & $8.32$  & $9.73$ & $10.74$  & $10.98$ &  $6.80$ & $7.78$ & $9.15$ & $9.79$ \\
       \hline
       \hline
	\end{tabular}
\end{table*}

The separation results on the MUSDB18-HQ test set are reported in Table~\ref{tab:test_results}. First, we compare our linear fader-based inference procedure to an OLA-based procedure, similar to that of the original BSRNN paper (see Section~\ref{sec:protocol-inference}). Varying hop sizes using this OLA procedure yields a 0.1 dB difference on the \texttt{vocals} track, which is consistent with the original results~\cite[Table II]{Luo2023bsrnn}. The OLA and fader techniques yield similar results on average, but using OLA with a 0.5~s hop size largely increases inference time by a factor of~6. This confirms the advantage of our linear fader for faster inference. Similar conclusions can be drawn using the optimized model.

We observe that our implementation falls behind the original test results~\cite{Luo2023bsrnn} by 0.5~dB on average. This motivates further refining our pipeline to better match these. Table~\ref{tab:test_results} also includes the performance of our optimized model denoted oBSRNN and described in Section~\ref{sec:results-opt}. This model is able to bridge the performance gap, as it largely improves our BSRNN implementation, and it even outperforms the original  paper's results by 0.6 and 1.2 dB in terms of uSDR and cSDR, respectively. This optimized model is therefore an interesting alternative to BSRNN, as it is openly available and yields better results.

\noindent \textbf{A note on SIMO-BSRNN} We investigated additional modeling variants that are implemented in SIMO-BSRNN~\cite{Luo2024simo}. This includes a finer band-split scheme, sharing the band-split and sequence and band modeling modules across sources, and a masker that uses the neighboring frequencies as context. For brevity, we do not present the corresponding experiments here, but details are provided in the supporting document. Our optimized version, denoted oBSRNN-SIMO in Table~\ref{tab:test_results}, largely outperforms SIMO-BSRNN and actually performs on par with BS-RoFormer (\textit{cf}. results from Table~\ref{tab:sdr_comparison}). Since this model reaches state-of-the-art MSS quality, one might consider using it if the goal is to achieve maximum performance.

\section{Discussion}
\label{sec:discussion}

This replication study has shown insightful in many regards. It notably extended the original BSRNN paper's analysis by investigating a variety of parameters such as architectural (e.g., the masker's MLP size), training-related (e.g., the contribution of each term in the loss), or data-related (e.g., the SAD preprocessing). Explicitly reporting such results, even though sometimes \textit{negative}, is beneficial to other researchers as it spare them a costly (re)investigation.
Besides, these detailed results pave the way for potential improvements in terms of model reduction. For instance, reducing the masker size or leveraging our naive stereo approach yielded a similarly performing but lighter \texttt{bass} model. Alternatively, using a small \texttt{drums} model with attention is both competitive with a larger variant and significantly lighter. Our experiments also illustrate that one cannot reuse \textit{as-is} building blocks from other papers. For instance, the dilated convolution~\cite{wang23ca_interspeech}, or the multi-head modules~\cite{Chen2024}, were shown effective in the corresponding papers, but they were used in conjunction with other architectural blocks and integrated in a different pipeline. The same applies to the variants incorporated in the SIMO-BSRNN model~\cite{Luo2024simo} where only the masker context is evaluated independently.

While our analysis mostly focused on performance results in terms of numerical values of SDRs, the fundamental reproducibility issues lie beyond this sole aspect, which echoes prior work on this topic in the field of music information retrieval~\cite{Six2018,McFee2019}. A first problem is that the discrepancy between results is currently unexplained, and can come from whether one or a combination of factors including, but not limited to, all variants considered in this study. Since no official implementation for the whole pipeline is available, there is no guarantee that we did not make any mistake or that there is not a substantial mismatch between implementations. A second problem is that the replication process itself is extremely time-consuming, which hampers research development. Indeed, the whole implementation of the project (including prototyping and debugging), as well as all extensive training to tune hyper-parameters, could be avoided if the code (along with training routine, hyper-parameters and eventually weights) was released. This in turn would promote faster research that builds upon this project, rather than re-implementing it.

The specific deep learning nature of this project brings additional reproducibility costs. Indeed, while deep learning research has a considerable environmental footprint and should move towards its systematic monitoring and reduction~\cite{Henderson2020, Strubell2020}, non-reproducible research exacerbates this issue, as outlined by our results in Section~\ref{sec:results-energy}. In all fairness, part of this energy cost is due to our own implementation errors, which resulted in, e.g., interrupted or redundant training runs, but we believe that most music/audio researchers are not machine learning or coding experts, therefore these pitfalls are likely common. In addition, our study exhibits energy and job ratios that are typical of similar projects' development phase. For instance the authors from a previous study~\cite{Strubell2020} report that the electricity cost of their whole project is about $2000$ times that of training a single model, and they report running almost $5000$ jobs in total, including many that crashed. Overall, such a project has a substantial energy cost, which should be systematically reported and taken into account when performing model selection. Besides, this study's footprint would have been substantially reduced upon availability of the code, documentation, and proper hyperparameter selection, as many trial-and-error experiments could have been avoided.

Finally, deep models' performance strongly depend on the hardware, as it affects the training protocol. On the one hand, given the impact of the (global) batch size on training speed and convergence, and thus on the final performance, accommodating the available hardware requires to consider and carefully fine-tune adaptation strategies (e.g., scaling the learning rate or accumulating gradients), which represents a substantial extra experimental burden. On the other hand, reproducibility might be plain impossible in practice for public institutions with limited computing capacity. For instance, as outlined in  Section~\ref{sec:motivation}, training a single BS-RoFormer source model~\cite{Lu2024} would take about 4 months if properly adjusted to fit our largest cluster, which prevents any in-depth experimental study. Such a discrepancy between hardware capacity sets restrictions on candidates for replication studies, which in turns should be taken into account when benchmarking a novel system against competing models in research studies.

\section{Conclusion}
\label{sec:conclu}

In this paper, we have addressed the issue of reproducibility in music separation research. We outlined several drawbacks of state-of-the-art MSS models, and we proposed to replicate and extensively analyze the BSRNN model. We implemented and released a fully functioning pipeline, including optimized models that perform on par with state-of-the-art systems. We also discussed the various costs associated with this project, notably in terms of energy consumption. This study resonates with works from other machine learning fields such as recommender systems~\cite{Dacrema2019} or language processing~\cite{Belz2021}, and could be relevant for other tasks such as speech enhancement and separation, which are prone to similar risks as MSS.

Beyond its focus on performance, the core contribution of this work is to recall that reproducibility is a fundamental aspect of the scientific endeavour. We encourage our colleagues to adopt open research practices, notably via releasing their code with proper documentation~\cite{McFee2019}, albeit we acknowledge that sometimes this is not possible due to copyright matters or company policy. In such cases, we respectfully suggest that the resulting papers should be considered mostly for their methodological or theoretical merits, and numerical performance results should be reported with extra care, especially when conducting model comparison. Such practices will foster a more transparent, reliable, and cost-effective research.

\section{Acknowledgements}
\label{sec:acknow}

All computation were carried out using the Grid5000 (https://www.grid5000.fr) testbed, supported by a French scientific interest group hosted by Inria and including CNRS, RENATER and several Universities as well as other organizations. We thank Jianwei Yu (author of the BSRNN paper) for answering some of our implementation-related questions. We also thank Christopher Landschoot for fruitful discussion related to his own unofficial BSRNN implementation, and Stefan Uhlich for insightful discussions on UMX and the MDX challenge, which were helpful in improving our work.

\appendices
\section{Reporting MSS models performance}
\label{sec:appx:perfreport}

This appendix describes the method for collecting the test results from Table~\ref{tab:sdr_comparison}. This allows one to avoid ambiguities or inconsistencies that might occur when cross-comparing papers.

Let us first recall that the MUSDB18 dataset described in Section~\ref{sec:protocol-data} comprises a high-quality version denoted MUSDB18-HQ~\cite{Rafii2019musdb18hq}, and a \emph{compressed} version, herein denoted ``nonHQ''~\cite{Rafii2017musdb}. The co-existence of these two versions, although beneficial to the community as it provides more flexibility, might complicate reporting.
\begin{itemize}
    \item In its original paper, the ResUNet model~\cite{Qiuqiang2021resunet} is evaluated on MUSDB18-nonHQ. Nevertheless, it is commonly reported in separation benchmarks along with other models tested on MUSDB18-HQ~\cite[Table 3]{Defossez2021hybrid}, which is somewhat confusing; and sometimes reported as computed on the HQ test set~\cite[Table 3]{Chen2024}, which is erroneous. We do not include this model in Table~\ref{tab:sdr_comparison}, since its performance on MUSDB18-HQ is not available.
    \item UMX~\cite{Stoter2019umx} official repository\footnote{\url{github.com/sigsep/open-unmix-pytorch}} reports results on both datasets. A follow-up paper reports performance on MUSDB18-nonHQ~\cite[Table 1]{Sawata2021xumx}, but the numbers correspond to the MUSDB18-HQ performance from the official repository. We do not include this model because of this confusion, as well as its overall low performance compared to more recent methods.
    \item Similarly, the original paper accompanying the KUIELab-MDX-Net model~\cite{Kim2021kuielab} reports performance on MUSDB18-nonHQ, while these same numbers are then reported as performance on MUSDB18-HQ in the HT Demucs paper~\cite{Rouard2023htdemucs}. Then, the BSRNN paper~\cite{Luo2023bsrnn} reports performance on both datasets, thus we extract the HQ results from this source to report it in Table~\ref{tab:sdr_comparison}.
    \item Lastly, we do not include other popular models such as Spleeter~\cite{Hennequin2020spleeter}, or D3Net~\cite{Takahashi2021d3net}, since to the best of our knowledge, their performance is solely reported on MUSDB18-nonHQ.
\end{itemize}

An additional source of confusion is the existence of several model variants. For instance, the Hybrid Demucs original companion paper~\cite{Defossez2021hybrid} reports the performance of both a basic model, and that of an optimized bag of models. Subsequent papers then report either one~\cite{Rouard2023htdemucs} or the other~\cite{Luo2023bsrnn,Lu2024}, without specifying which one explicitly nor providing justification.  In Table~\ref{tab:sdr_comparison} we report results that correspond to the optimized bag of models, since it yields the largest SDR. The same applies to HT Demucs, for which we report the best results from the original paper~\cite{Rouard2023htdemucs}, except for the average SDR of the model trained with extra data (we replace the $9.20$ value with the actual average of $9.27$ dB).

Finally, the results corresponding to the remaining models (CWS-PResUNet, TFC-TDF UNet v3, and BSRNN and variants) are reported from the related original publications.

This description acts as a reminder of the importance of reporting results from other papers with extra care. In particular, we insist on disclosing which version of the dataset is used, as well as which model variant is selected. This ensures proper and fair comparison, especially for model ranking.

\bibliographystyle{IEEEtran}
\bibliography{references}

\end{document}